\documentclass[epjCONF]{svjour}
\usepackage{graphics}
\usepackage[varg]{txfonts} 
\usepackage[latin1]{inputenc}

\session-title{A Universe of Dwarf galaxies:}
\begin{document}
\title{Kinematically detected polar rings/disks in blue compact dwarf galaxies}
\author{A. Moiseev\inst{1}\fnmsep\thanks{\email{moisav@gmail.com}}}
\institute{Special Astrophysical Observatory, Nizhnii Arkhyz, 369167 Russia}
\abstract{
Polar ring galaxies are systems with nearly orthogonally rotated components. We have found the gas on polar (or strongly inclined) orbits in two BCD galaxies using ionized gas velocity fields taken with a Fabry-Perot interferometer  of the SAO RAS 6-m telescope. Our analysis shows that all ionized gas in Mrk 33 is concentrated in a compact disk (3 kpc in diameter) which rotates in the polar plane relative to the main stellar body. The gaseous disk in Mrk 370 has a more complex structure with a heavily warped innermost part. The presence of polar gaseous structures supports an idea that current the burst of star formation in these galaxies is due to the external gas accretion or merging. A possible fraction of polar structures among BCD galaxies seems to be very large (up to 10-15\%).
} 
\maketitle
\section{Observations}

The observations were made  with a scanning Fabry-Perot interferometer (FPI) mounted within the
multimode instrument SCORPIO \cite{Afanasiev2005} at the prime focus of  the Russian 6-m telescope. The spectral range was equal to $13\AA$ and the spectral resolution was $0.8\AA$ (about 35 km/s) for a $0.36\AA$ sampling. The spectral interval was centered on the redshifted emission lines [NII]$\lambda$6583 (Mrk 33) and H$\alpha$ (Mrk 370). The observational data were reduced using the IDL-based software package \cite{MoiseevEgorov2008}. The final spatial resolution was about $2-2.5''$. We fitted emission line profiles to the Voigt function, the profile fitting results were used to construct the two-dimensional fields of the line-of-sight velocities of ionized gas, maps of velocity dispersion, and images in the corresponding emission lines.

\section{Mrk 33}

It is a blue compact galaxy with structural properties corresponding to a dust-lane dE \cite{Bravo-Alfaro2004}. Perosian et al.\cite{Petrosian2002} have studied its circumnuclear  ($r<5''$) gas kinematics and concluded  that the position angle ($PA$) of the kinematic major axis significantly differs  from the photometric major axis of the galaxy. Our observational data allows us to study the ionized gas kinematics at   larger distances from the nucleus - the [NII] emission was detected up to $r=15''$. The velocity field of the ionized gas can be approximated by the model of a circular rotating thin disk. The major axis of this disk  differs significantly from the major axis of the outer isophotes. The major axis position angle   and inclination  derived from the ionized gas velocity field are $PA=163\pm4^\circ$  and $i=47\pm5^\circ$, respectively. For the outer oblate spheroid we adopted the orientation parameters based on photometry analysis \cite{Amorin2009}:
$PA=116^\circ$ , $i=59^\circ$. These values imply two formal solutions for the relative inclination between the ionized gas disk and the host galaxy main plane:  $\Delta i=85\pm6^\circ$ or $39\pm6^\circ$. The former  corresponds to the stable polar structure. Based on the low-resolution kinematic data in HI,  Bravo-Alfaro et al. \cite{Bravo-Alfaro2004} also suggested that Mrk~33 has recently captured gas in a close interaction or merger with an initially gas-rich companion. Our observations show that the inner polar ionized gas disk has a similar kinematics with the external HI structure. The orientation of the inner ($r<10''$) optical isophotes is in a good agreement with the ionized gas disk position. This fact indicates that a significant fraction of stars was already formed from the gas on  polar orbits, therefore the inner gaseous disk was stable for least several dynamical times
($\sim10$ Myr).

\section{Mrk 370}

\begin{figure*}
\resizebox{2\columnwidth}{!}{
\includegraphics{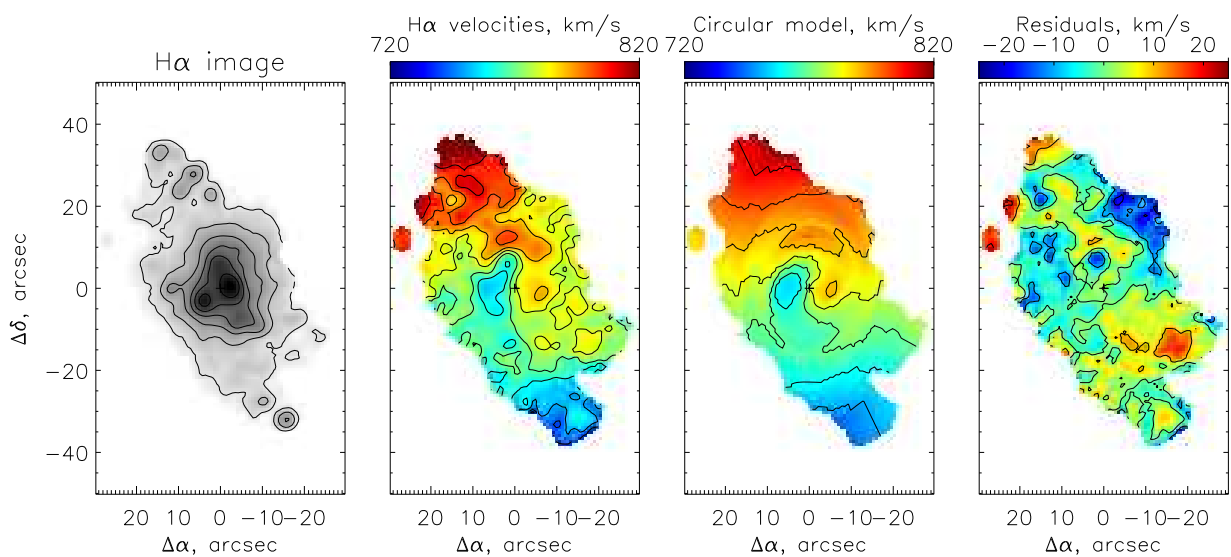}
}
\caption{Mrk 370, the maps derived from FPI data cube: H$\alpha$ monochromatic image, velocity
field of the ionized gas, a titled-ring model of the circular rotation of the warped disk and  the residual velocity
map (observations minus model).}
\label{Moiseev_fig1}
\end{figure*}

Cairos et al. \cite{Cairos2002} found that the current star formation with the age of 3-6 Myr took place in numerous knots in
this luminous ($M_B=-17.20$) BCD galaxy. Our FPI data show that external emission knots rotate in a plane
roughly coinciding with the stellar disk of the galaxy. At smaller radii ($r<700-800$ pc), circular orbits in the ionized gas
disk change  orientation abruptly (see Fig.~\ref{Moiseev_fig1}). The intrinsic orientation of the disk changes through  $\Delta i=55-70^\circ$  (it depends on the accepted outer disk inclination). We suppose that a coherent warped disk observed in Mrk 370 is a small-scale analogue of strongly twisted disks recently found in the galaxies NGC 2685 \cite{Josza2009} and NGC 3718 \cite{Sparke2009}.

\section{Summary}

In both studied  galaxies the ionized gas in the circumnuclear region ($r<1-2$ kpc) rotates in the plane which is orthogonal  (or significantly inclined) to their stellar disks. Similar circumnuclear polar disks where recently found  in more than 30 nearby galaxies (see \cite{Moiseev2010} for referefences and discussion). Though a detailed process of the inner polar disk formation is still an open question \cite{Moiseev2010}, the most likely formation mechanism is the same with classical polar ring galaxies (PRG) --  merging or accretion of external gas clouds with a specific direction of orbital momentum  \cite{BournaudCombes2003}.  It is considered that  PRGs  are  rare objects -- according  to \cite{Whitmore90}, only  $0.5-5$\% of S0 galaxies possess polar rings and this percentage  is significantly smaller for late-type galaxies. However, a  possible fraction of polar structures among BCD galaxies seems to be larger (up to 10-15\%). For instance, 5 galaxies out of 28 nearby luminous BCDG in the sample by Cair{\'o}s et al. \cite{Cairos2001} possess polar structures: Mrk 33, Mrk 370, Mrk 314 \cite{Shalyapina2004}, II Zw 71 \cite{Cox2001} and III Zw 102 \cite{Moiseev2008}.  Therefore the current burst of star formation in these galaxies can be connect with the same interaction event (accretion or merging)  which also  transported  gas clouds on polar  orbits.

\begin{acknowledgement}
Based on the observations collected with the 6-m telescope of the Special Astrophysical Observatory of the Russian Academy of Sciences, which is
operated under the financial support of the Science Department of Russia (registration number 01-43).
This work was supported by the Russian Foundation for Basic Research (projects no.~09-02-00870) and  by the `Dynasty' Fund.
I am very grateful to the Organizing Committee for their support of my stay  in Lyon during this Meeting.
\end{acknowledgement}


\end{document}